\documentclass{webofc}

\usepackage[varg]{txfonts}   
\usepackage{hyperref}
\usepackage{url}
\hypersetup{colorlinks=true,citecolor=blue,urlcolor=blue,linkcolor=blue}
\begin{document}
\title{An End-to-End Generative Diffusion Model for Heavy-Ion Collisions}
\author{\firstname{Jing-An} \lastname{Sun}\inst{1,2} \and
        \firstname{Li} \lastname{Yan}\inst{1,3}\fnsep\thanks{\email{cliyan@fudan.edu.cn}} \and
        \firstname{Charles} \lastname{Gale}\inst{2} \and
        \firstname{Sangyong} \lastname{Jeon}\inst{2}
}

\institute{Institute of Modern Physics, Fudan University, Shanghai, 200433, China 
\and
           Department of Physics, McGill University, Montreal, Quebec H3A 2T8, Canada 
\and
           Key Laboratory of Nuclear Physics and Ion-beam Application (MOE), Fudan University, Shanghai 200433, China
          }

\abstract{Heavy-ion collision physics has entered the high precision era, demanding theoretical models capable of generating huge statistics to compare with experimental data. However, traditional hybrid models, which combine hydrodynamics and hadronic transport, are computationally intensive, creating a significant bottleneck. In this work, we introduce DiffHIC, an end-to-end generative diffusion model, to emulate ultra-relativistic heavy-ion collisions. The model takes initial entropy density profiles and transport coefficients as input and directly generates two-dimensional final-state particle spectra. Our results demonstrate that DiffHIC achieves a computational speedup of approximately $10^5$ against traditional simulations, while accurately reproducing a wide range of physical observables, including integrated and differential anisotropic flow, multi-particle correlations, and momentum fluctuations. This framework provides a powerful and efficient tool for phenomenological studies in the high-precision era of heavy-ion physics.
}
\maketitle

\section{Introduction}
The primary goal of ultra-relativistic heavy-ion collisions conducted at the LHC and RHIC is to create and study the quark-gluon plasma (QGP), a state of matter where quarks and gluons are deconfined~\cite{Shuryak:2014zxa,Heinz:2013th}. Hydrodynamic models~\cite{Yan:2017ivm,Shen:2020mgh,hydroSchenke:2010rr,hydroSchenke:2010nt,hydroPaquet:2015lta,hydroGale:2021emg} have been remarkably successful in describing the collective, fluid-like behavior of the QGP. State-of-the-art simulations typically employ a hybrid approach, coupling initial state models (e.g., TRENTO)
with viscous hydrodynamic evolution (e.g., MUSIC)
, followed by particlization and hadronic transport (e.g., UrQMD).

Despite their success, these traditional models face a severe computational challenge. A single collision event simulation can take approximately 120 minutes on a single CPU core. This makes it computationally prohibitive to achieve the large statistics ($10^9-10^{10}$ events) required to study the hyperfine structure of QGP or perform high-precision analyses, such as nuclear structure effects, or the QGP's equation of state.

To overcome this limitation, we leverage machine learning, specifically generative artificial intelligence. We present the first application of a conditional diffusion model~\cite{Diffho2020denoising,Diffsohl2015deep,Diffsong2020score}, named DiffHIC (Diffusion Model for Heavy-Ion Collisions), as a fast, end-to-end surrogate model for heavy-ion collision simulations. This approach aims to learn the complex mapping from initial conditions to final particle spectra, a process governed by non-linear hydrodynamic and Boltzmann transport equations.

\begin{figure}[h!]
        \centering
        \includegraphics[width=0.5\textwidth]{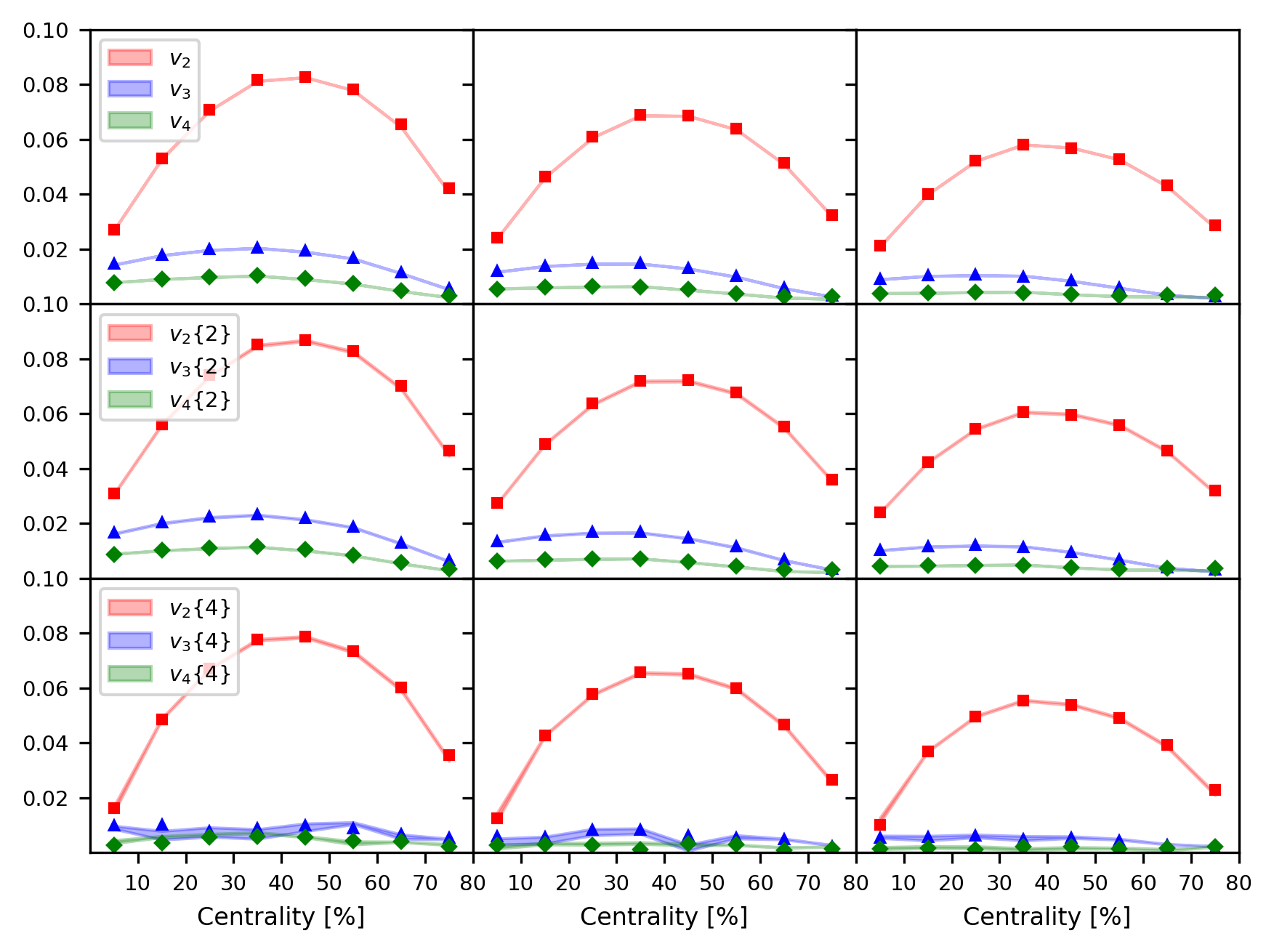} 
        \caption{Centrality dependence of integrated anisotropic flow. Filled symbols represent the ground truth from the hybrid model, while colored bands show the results from DiffHIC. The agreement is excellent across different orders ($n$), cumulants, and shear viscosities.}
        \label{fig:flow}
\end{figure}
\vspace{-1cm}

\section{Methodology}

\subsection{The Generative Diffusion Model}
Diffusion models~\cite{Diffsong2020score} are a class of generative models that learn a data distribution by reversing a gradual noising process. The process is defined by a forward stochastic differential equation (SDE) that transforms data into a simple prior (e.g., Gaussian noise), and a corresponding reverse-time SDE that transforms the prior back into the data distribution. The key to the reverse process is learning the score function, $\nabla_x\log p_t(x)$, which can be approximated by a neural network trained via denoising score matching.

Our model, DiffHIC, is a conditional diffusion model. It is trained to generate the final charged particle momentum spectra $\mathbf{S}$ conditioned on the initial entropy density profile $\mathbf{I}$ and the shear viscosity to entropy density ratio, $\eta/s$. The core of DiffHIC is a U-Net neural network architecture, which is widely adept in image-related tasks. The initial profile, conditioning parameters ($\eta/s$), and the diffusion timestep $t$ are fed into the network to predict the noise added at that step. Once trained, the model can generate new particle spectra by solving the reverse ODE starting from random noise, a process that is both deterministic and computationally efficient.

\subsection{Training and Validation Datasets}

To train DiffHIC, we generated a large dataset of Pb-Pb collisions at $\sqrt{s_{NN}} = 5.02$ TeV using a well-established hybrid model. The initial entropy density profiles were generated using the TRENTO model. The subsequent (2+1)D viscous hydrodynamic evolution was simulated with MUSIC, followed by particlization via the Cooper-Frye formula and hadronic rescattering with UrQMD. We generated 12,000 minimum bias events for each of three distinct values of shear viscosity: $\eta/s = 0.0, 0.1$, and $0.2$. For validation, an additional 10,000 events per $\eta/s$ value were generated across all centralities.

\section{Results and Performance}
\subsection{Computational Efficiency}
The most significant advantage of DiffHIC is its computational efficiency. Generating a single event with DiffHIC takes approximately 0.1 seconds on a single NVIDIA GeForce GTX 4090 GPU. This represents a speedup factor of roughly $10^5$ compared to the $\sim$120 minutes required by the traditional hybrid model on a single CPU. This acceleration transforms previously infeasible large-scale statistical studies into routine computations.

\begin{figure}
        \centering
        \includegraphics[width=0.6\linewidth]{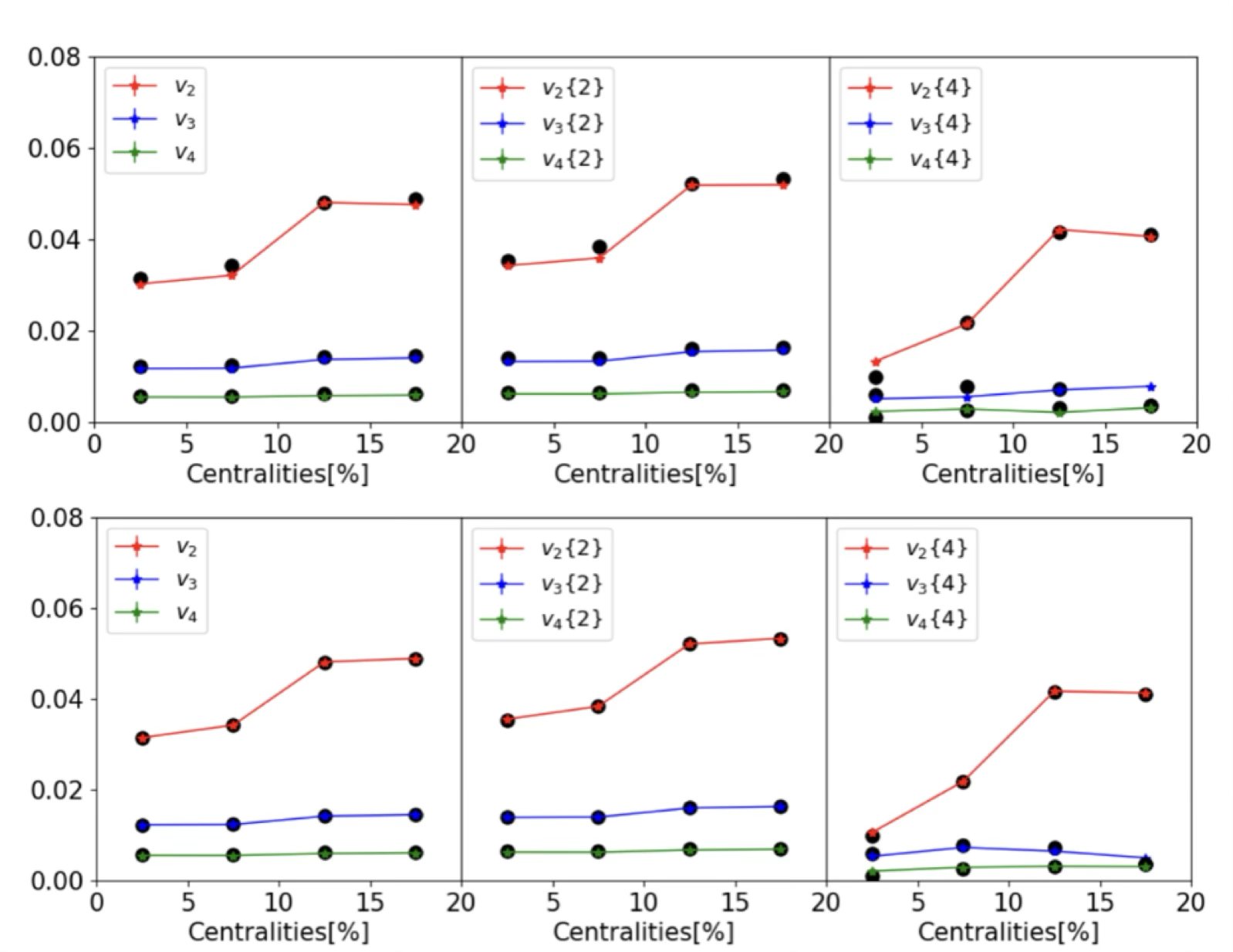}
        \caption{Symbols represent results from a hydrodynamic model, while lines show the output of a Generative AI model. The top row shows the pre-trained model, and the bottom row shows the improved results after fine-tuning with 500 new events.}
        \label{fig:finetune}
\end{figure}

\subsection{Physics Fidelity}
To validate the physical accuracy of DiffHIC, we compared its generated observables with the ground truth from the traditional hybrid model.

\begin{itemize}
    \item \textbf{Anisotropic Flow:} Anisotropic flow coefficients, $v_n$, are crucial for characterizing the collective dynamics. As shown in Fig.~\ref{fig:flow}, DiffHIC accurately reproduces the centrality dependence of integrated flow ($v_2, v_3, v_4$) calculated from both the event-plane method and multi-particle cumulants ($v_n\{2\}, v_n\{4\}$). The colored bands (DiffHIC) show excellent agreement with the filled symbols (ground truth) across all centralities and $\eta/s$ values.

    \item \textbf{Comprehensive Comparison:} We conducted a comprehensive comparison across a wide range of observables, including pT-dependent flow, mean transverse momentum fluctuations, and mixed harmonic correlations. As presented in Ref.~\cite{sun2024endtoendgenerativediffusionmodel}, the ratio of the observables generated by DiffHIC to the ground truth values are consistently close to unity, indicating the model's high fidelity. 
    
    \item \textbf{Flexibility and Fine-Tuning:} A key feature of our approach is its adaptability. We demonstrated that a pre-trained DiffHIC model can be quickly fine-tuned with a small dataset (e.g., 500 events) to incorporate new physics, such as the effects of nuclear deformation in Uranium-Uranium collisions, as shown in Figure~\ref{fig:finetune}. This highlights the model's potential as a flexible tool for exploring a wide parameter space without retraining from scratch.
\end{itemize}

\section{Conclusion and Outlook}

We have developed an end-to-end generative diffusion model, DiffHIC, for simulating heavy-ion collisions. This model achieves a remarkable $10^5$-fold speedup over traditional hybrid simulations while maintaining high physical fidelity across a lot of observables. DiffHIC effectively learns the intricate physics of QGP evolution and hadronization, providing a powerful and efficient framework for phenomenological studies.

Future work will focus on extending the model's capabilities. This includes incorporating more physical parameters (e.g., bulk viscosity, equation of state), moving towards 3D hydrodynamic simulations, and generating full particle clouds for greater kinematic detail. The success of DiffHIC paves the way for a new generation of fast, AI-driven simulators to meet the challenges of the high-precision era in heavy-ion physics.

\bibliography{reference}

\end{document}